# MyProLang – My Programming Language
A Template-Driven Automatic Natural Programming Language

Youssef A. Bassil, Aziz M. Barbar

*Abstract*—Modern computer programming languages are governed by complex syntactic rules. They are unlike natural languages; they require extensive manual work and a significant amount of learning and practicing for an individual to become skilled at and to write correct programs. Computer programming is a difficult, complicated, unfamiliar, non-automated, and a challenging discipline for everyone; especially, for students, new programmers and end-users.

This paper proposes a new programming language and an environment for writing computer applications based on source-code generation. It is mainly a template-driven automatic natural imperative programming language called MyProLang. It harnesses GUI templates to generate proprietary natural language source-code, instead of having computer programmers write the code manually. MyProLang is a blend of five elements. A proprietary natural programming language with unsophisticated grammatical rules and expressive syntax; automation templates that automate the generation of instructions and thereby minimizing the learning and training time; an NLG engine to generate natural instructions; a source-to-source compiler that analyzes, parses, and build executables; and an ergonomic IDE that houses diverse functions whose role is to simplify the software development process.

MyProLang is expected to make programming open to everyone including students, programmers and end-users. In that sense, anyone can start programming systematically, in an automated manner and in natural language; without wasting time in learning how to formulate instructions and arrange expressions, without putting up with unfamiliar structures and symbols, and without being annoyed by syntax errors. In the long run, this increases the productivity, quality and time-to-market in software development.

Future research can improve upon MyProLang so much so that it can be internationalized to support more natural languages. It can also benefit from OOP capabilities and from reusable programming libraries.

*Index Terms*—Automatic Programming Language, Natural Language Generation, Source-Code Generation, Source-to-Source Compiler.

## I. INTRODUCTION

Although today's programming languages are well observed and used worldwide, they are still syntactically sophisticated [1] since they contain confusing structures, complex expressions and uncommon symbols. Furthermore, the extra details and notations related to the formality and design of the language, like include-directives, compiler primitives, classes declaration, functions prototype and initialization statements, must be handled explicitly by the programmer, albeit they are of no use to his algorithm. Such intricacies and extra vestigial complexities prevent the language from being a good learning instrument, hamper the programming practice, and increase the probability of compilation and logical errors. Additionally, programming languages are inexpressive in that they don't use human natural language to represent codes and instructions. Naturalness of the language makes it possible for people to express their ideas in the same way they think about them, [2] this eases the process of implementing a mental plan into code. Finally, programming is not an automatic generative process, it completely relies on the programmer to manually construct the source-code word by word and instruction by instruction. This would consume time and slow down the development procedure.

As a result, we can admit that modern programming languages are not that easy to learn and manage, they bring in a lot of complications, they are erroneous, they don't reflect the programmer's intention, they are still manual, and they impose barriers for beginners and average people with no previous training and experience.

What we are proposing is this paper is a programming language characterized by its syntax simplicity and expressiveness, its ease to learn and use, its ability to automate programming processes, and its facility for rapid application development (RAD). Ultimately, programming would become a straightforward task, open to everyone in a way that end-users would simply become programmers.

## II. TERMINOLOGY

In this section, definitions of some concepts and key subjects that are in some way connected to our work, are given.

### A. Computer Programming

Programming is the process of building executable applications from the arrangement of certain words and symbols called instructions. This arrangement is specified by a set of grammatical rules and other rules indicated by the programming language itself [3].

At early stages, the programmer has first to type instructions in a text editor, and then he has to compile the source-code to check for syntax errors, and generate machine code that can be linked and executed as a standalone application on a specific platform.

### B. Source-to-Source Compiler

A compiler is usually another computer program whose job is to scan and parse programmer's source-code to capture syntax errors, and eventually generate machine code understandable by a particular machine-native instruction set [4].

A source-to-source compiler is a type of compilers that takes a high-level language as input and produces another high-level language as output. Such type of compilers should operate upon another traditional compiler whose purpose is to convey higher abstractions into lower native code.

### C. Source-Code Generation

Source code generation is the process of generating programming source-code through any of various means such as XML schemas, graphical templates or UML models. Such means are the input for the code generator, and the spawned source-code is the output. Code generators are usually separate applications or they can be embedded into development-environment tools.

GUI Template-driven source-code generators are a type of code generators that guide the user through an automated process of pop-up graphical windows or forms in order to generate a predefined piece of code. For instance, the query by example (QBE) system helps you create your own SQL queries by selecting combo boxes, and filling up text fields without manually writing any line of code.

### D. Natural Language Generation

Natural Language Generation (NLG) is one of the emerging fields in natural language processing. An NLG system converts non-linguistic data, such as graphical models or templates, into natural linguistic information such as an English text [5]. An NLG source-code generator is a system that generates natural source-code pulled out from a certain form of input.

One of the central goals of NLG is to investigate how computer programs can be made to produce high-quality, expressive, uncomplicated and natural language text from computer-internal sophisticated representations of information. An NLG system can be defined as follows [6]:

*Goal*: Computer software which produces understandable and appropriate text in English or other human languages.

*Input*: Some underlying non-linguistic representation of information.

*Output*: Documents, reports, explanations, help messages, and other kinds of texts.

*Knowledge sources required*: Knowledge of the domain of the system being constructed.

### E. Automatic Programming Languages

Automatic programming languages are types of computer languages that generate the coding for you. They are built over a source-code generator, which uses a variety of gears such as templates or models, to extract specifications from the user, and to ultimately produce the desired instructions and code. These generated instructions can be represented in a well-known programming language such as C++, or it can be represented in a proprietary language and thus a proprietary compiler should be built to do the translation to lower layers.

An automatic programming language can exploit natural language generation techniques with the purpose of producing natural source-code instead of the 3GL source-code.

### III. PREVIOUS WORK

A lot of automatic programming languages and code generators exist nowadays. However, in this section, we are mostly concerned with known and acknowledged automatic programming languages oriented to IBM compatible PCs.

QBE, otherwise known as query by example, is a form of automatic programming language. It was developed by Moshé M. Zloof at IBM Research during the mid 1970s. QBE is a database automatic query language for relational databases. It is composed of graphical tables through which the user would select table names, choose elements and set search conditions. The QBE system would then generate and execute the appropriate SQL statement [7]. This approach allows the user to carry out powerful searches without the need of having to learn the formal syntax of SQL.

Microsoft Corporation in the early 1990's introduced the MS Visual Studio IDE to support a group of programming languages. MS Visual Studio allows the designing of graphical user interfaces in an interactive manner, while the compiler, invisibly, generates the corresponding source-code. In MS Visual Studio 2005 ADO.NET connections and objects can be created via templates while the environment generates the corresponding source-code, without the programmer's knowledge.

Snippet is yet another form of automatic programming that was introduced by Microsoft Corporation. They are a collection of predefined and reusable source-codes inserted by the programmer in the current document. Examples of Snippets could be if-blocks and for-loops structures. Their purpose is to relieve the programmer from the burden of memorizing certain structures. With Snippets, programmers can easily drop in code the same way they "copy-paste" text.

Macros are also another common form of source-code generation. Macros usually hide numerous and complex instructions by a single program statement, thus making the programming task more comprehensive and error-free. Almost all Assembly languages implement a macro processor that replaces patterns in source-code according to relatively simple rules.

### IV. PROJECT MOTIVATIONS

The chief problems that are tackled in this paper are described herein.

***Syntax Complexity and Programming Errors***: All the programming languages available today on the market have their own syntax and grammar. Usually such syntax is not clear-cut since it has special format and rules, and it contains weird symbols and unusual notations. This poses barriers which, in turn complicates the programming process. Moreover, it ignites dozens of compilations and syntax errors that make the language error-prone and subject to bugs and flaws. For instance, to create a string of characters in C, you may write *char\* name="John";*. Notice the meaningless use of \*, the double quotes and the semicolon. Another form of complexity can be manifested in the subtle distinctions in syntax that lead to confusion and non-consistency [8]. For instance, in C, there are three different kinds of braces used in various situations: *{}, ()* and *[]*. In the Java language, all control structures do not end with a semicolon except for the *do-while*. Also in C, indexing of arrays is done with square brackets; however, initialization is done with curly braces. As a result, novices get confused when a single notation can accomplish different effects [9].

*Transparency and Abstraction:* In traditional programming languages, low-level details and unnecessary terms must be handled by the programmer and not by the language itself. For instance, the compiler directives and primitives, the superfluous keywords and symbols, must be carried out by the programmer even though it does not serve and it is not related to his logical algorithm. For instance in Java, to just print on the screen the message "hello world" you still require a bunch of additional keywords and statements.

```
public class TestApp {
    public static void main(String[] args)
    {
        System.out.println("Hello World") ;
    }
}
```

*Syntax Inexpressiveness*: In available programming languages, it is impossible to express your program's source-code in a human natural English-like sentence. The odd format of expressions and instructions in the language would increase the gap between the programmer's natural plan and the practical implementation. As a design principle for all learning environments, the translation process from the student's internal plans to the solution's external representation should be minimized [10]. It is more familiar to say *print hello on the screen* then saying *System.out.println("Hello") ;*.

*Non-automation of Programming Proceedings*: Nearly all programming instructions are written manually by the programmer through traditional ways. Every instruction and declaration must be typed literally in a text editor through the machine keyboard. This involves time, effort and reserves resources.

*Language Learn Ability*: Mixed together, all the previous problems would hinder people from learning programming effortlessly. In other words, gaining knowledge of the language would not be that easy and would require days or even months to become skilled at. Yet it would constitute an obstacle and a nightmare for the new learners and the new programmers.

Our incentive is to provide an uncomplicated, transparent, less error-prone, natural, automatic, and an easy to learn programming language, yet has the ability to develop general-purpose computer programs.

V. PROPOSED SOLUTION

Our proposed solution is a template-driven automatic natural programming language used to develop computer programs. It is MyProLang which allows you to build computer applications based on source-code generation. It utilizes graphical templates or GUI forms, populated interactively with specifications data by the programmer, to generate natural proprietary-language source-code. This source-code can be eventually compiled into executables.

MyProLang is a combination of a proprietary programming language; a set of automation templates to automate programming procedures; an NLG engine to convert user specifications from templates into natural text; a source-to-source compiler to compile and generate executables; and an IDE to better interface the programmer with the machine. Fig. 1 depicts the five logical modules or chunks of MyProLang.

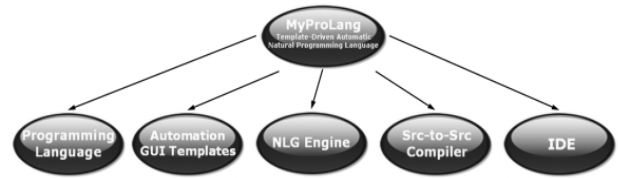

Fig. 1 - Logical Modules

*A. Programming Language*

MyProLang is a general-purpose imperative procedural, statically typed, type safe and compiled natural programming language oriented to IBM compatible PCs. It is used to write general-purpose standalone computer executables. It embraces a set of concepts that exist in any modern procedural language. For instance, MyProLang includes declaration of variables with different data types (*Number* and *String*). It also provides a one-dimensional static array structure. In addition, it supports arithmetic and logical calculations and provides relational operations. Its different operators can be summarized as follows: +,-,*, /, %,(), And, Or, Greater, Smaller, Greater or Equal, Smaller or Equal, Equal, Not Equal and the assignment operator =. Furthermore, the ampersand operator & is used for string concatenation. MyProLang also supports different control structures such as IF, IF-ELSE and IF-ELSEIF, REPEAT, SELECT with block compound statements and multiple operands condition statements. Moreover, MyProLang provides an input and output dialog boxes to read in data from the keyboard, and to display information on the screen.

*B. Automation Templates*

The automation templates play a leading role in our solution. They are used to systematically guide the programmer while generating instructions. MyProLang's automation templates are GUI based, in that they are composed of GUI elements such as text fields, combo boxes, list views and buttons. For each type of instruction there exits an appropriate template that the programmer opens, sets specification data such as the name of a variable, its type and its initial value or the message to be printed on the screen, and then, with just a button click, the proper instruction would be generated. This automated approach relieves programmers from hand-writing their source-code. In that way, programmers don't have to bother themselves with putting together expressions and arranging keywords and identifiers in the proper and correct order. They have just to follow the graphical templates to generate instructions. More to the point, automation templates provide a transparent layer that shields the programmer from the sophisticated low-level inner structure of the syntax. In other words, programmers don't have to care anymore about memorizing and learning terms and notations since everything is mimicked by the automation process. Besides, automation promotes RAD discipline by speeding application development, and by focusing on building applications in a very short period of time.

*C. NLG Engine*

The NLG engine is considered a Template Filling NLG system type because it generates programming instructions in English natural language through filling variable slots in a template primarily represented by a graphical template. MyProLang's NLG engine can be described as follows:
**Goal**: Generator of natural source-code
**Input**: GUI Templates populated with data by the programmer.
**Output**: Natural programming source-code.
**Domain**: General-purpose computer programming.

Fig. 2 shows the MyProLang's NLG system with its major components.

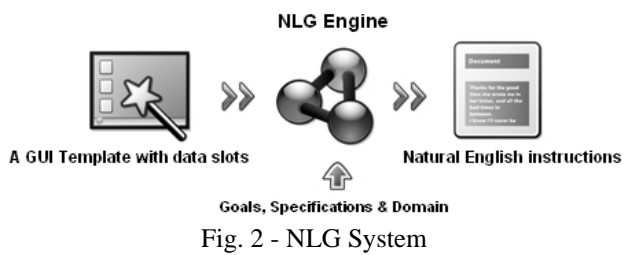

Fig. 2 - NLG System

*D. Source-to-Source Compiler*

MyProLang's compiler is a source-to-source compiler which converts specifications from templates into a high-level language mainly C#. For this reason another off-the-shelf compiler is needed to do the conversion to lower level. MyProLang's source-to-source compiler is composed of a lexical analyzer, recursive-descent parser of context-free grammar, a semantic analyzer and a code generator. At the beginning it validates the data supplied into the templates against the predefined language rules e.g. if a variable being declared already exists or if an assignment is incompatible with the variable data type or if an identifier contains illegal characters etc. Then it scans and parses several input fields to produce tokens and verify that no grammatical rules are violated. For instance correctness of arithmetic expressions, relevant arrangement of string concatenation operators and structure of logical conditions are all examined at this stage. Next, it internally produces an equivalent intermediate language representated in C#.NET 2.0. The NLG engine now runs and generates the corresponding natural source-code. This natural code is the only code visible to the programmer. The inner intermediate C# code produced previously, is then send to a C# compiler whose job is to convert the C# code generated by our source-to-source compiler into MS .NET bytecode also called Microsoft Intermediate Language (MSIL) [11], [12]. Eventually they will be transformed into an executable application by the .NET linker. Once completed, the produced executable file can be run on top of any .NET 2.0 Common Language Runtime (CLR). Fig. 3 depicts the MyProLang process flow diagram representing the different stages and the corresponding actions.

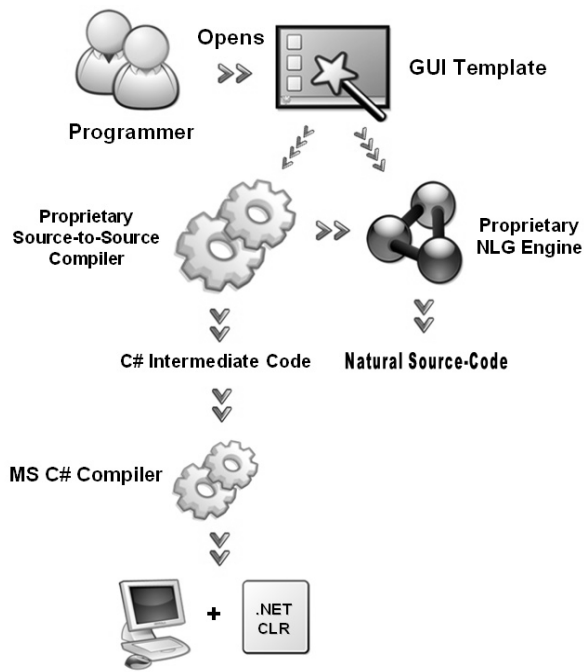

Fig. 3 - Process Flow

To make things clearer, a series of steps that show the flow of MyProLang processes, starting from opening a template till the production of an executable program, are written

**Step 1:** A Programmer opens a particular template with the purpose of generating a particular instruction.
**Step 2:** The Programmer follows the template and fills-in the required fields.
**Step 3:** Once done, the programmer clicks on the **generate** button to generate the instruction, and finally dismiss the template.
**Step 4:** MyProLang's source-to-source compiler extracts the populated data from GUI elements. It then scans, parses and warns of any errors found. Finally, it generates, internally, an intermediate language represented in C#.
**Step 5:** MyProLang's NLG engine then generates the natural source-code represented in our proprietary language. The output of the NLG engine is the only code visible to the programmer.
**Step 6:** When the programmer clicks on **compile**, in the main IDE toolbar, he is in fact only calling the off-the-shelf C#.NET compiler which reads the C#.NET instructions produced in step 4 from an internal buffer, and then generates .NET bytescode to eventually build the executable application.

*E. Integrated Development Environment*

MyProLang's IDE is the major container of all other components. It is characterized by a friendly, ergonomic and intuitive programming interface. It is fully integrated with GUI items, automation templates and houses the source-to-source compiler and the NLG engine. Besides, it features a set of tools that promote extensibility, storage and multimedia such as saving source-code on local disk and printing it. It also supports C# and Java translators, and a voice assistant tool that helps pin point compilation errors.

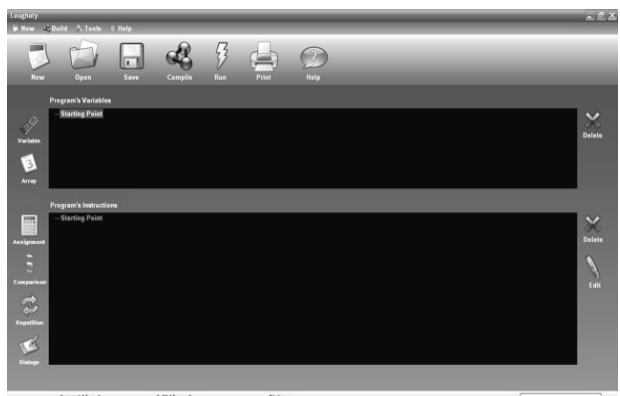

Fig. 4 - Main panel of the IDE

Fig. 4 illustrates the main development environment and interface of MyProLang. It is composed of two placeholders, one for global variables generation, and the other for instructions generation, a toolbar along with a menu bar that contains a set of functionalities such as saving, printing and compiling, and a left panel that contains a set of graphical icons used to launch templates, necessary to generate instructions.

## VI. IMPLEMENTATION AND TESTING

MyProLang is implemented in MS C#.NET 2005 using MS Visual Studio 2005 under the MS .NET Framework 2.0, with more than 20000 lines of code.

A demonstration of a sample test that exhibits how to generate a variable declaration is presented. The variable is called *Radius,* it is of type *Number*, and contains the value *25*. Fig. 5 shows the particular template for declaring a variable. Fig. 6 shows the generated natural variable declaration instruction. Fig. 7 depicts a complete generated source-code program for finding the average of three numbers.

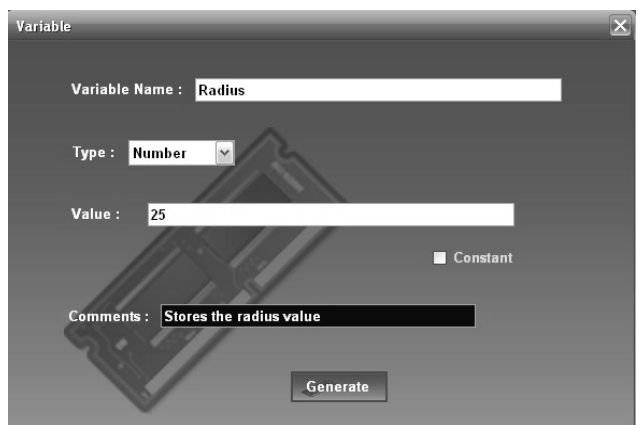

Fig. 5 - Automation template for generating a variable declaration

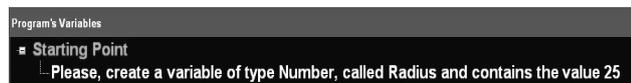

Fig. 6 - Natural source-code generated after dismissing the template

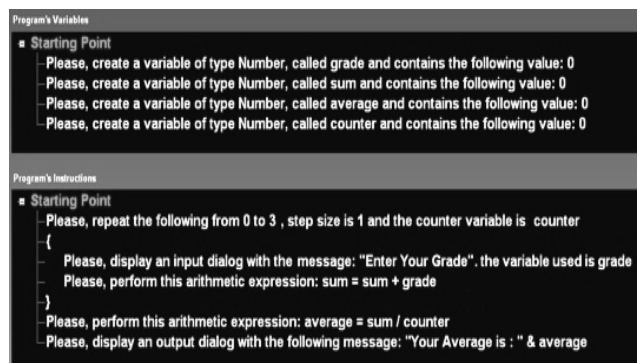

Fig. 7 - Complete generated program for finding the average of three numbers

## VII. RESULTS

The MyProLang presented a distinctive approach to help ease, simplify, familiarize, and automate the writing of computer programs.

A proprietary natural programming language based on unsophisticated syntax and grammatical rules was presented. In addition, to a programming environment that leverages the concept of automatic programming through graphical templates and natural source-code generation. It is a complete framework for producing general-purpose computer applications with just few mouse clicks.

The problem of syntax complexity and programming errors were solved by designing a plain natural programming language, which is undemanding, and easy to learn and use. It features a very expressive and familiar way to construct source-codes.

Automation templates automated programming procedures by substituting manual coding with the automatic generation of source-code. Besides, it also shielded the programmer from arranging and putting words together to mold instructions. Hence, programmers have no need to learn in depth and dig into the basic language specifications, they have just to follow intuitive templates, and the rest is automatically handled, implicitly, by the computer. Additionally, it accelerated the development time and promoted RAD practice.

## VIII. CONCLUSIONS

MyProLang is an open programming language for all, including end-users, students, junior developers and expert programmers. Consequently, anyone can program and build programs in a clear, familiar, systematic and rapid manner without dealing with ambiguous notations and bizarre symbols, without spending time in learning rules and formal structures, and without becoming frustrated by syntax errors and programming hazards.

On the long run, MyProLang would increase the productivity, quality, and time-to-market in software development.

## IX. FUTURE WORK

As further research, programming key technologies are to be added to the language such as classes, objects and other OOP features. Also a reusable set of built-in libraries beneficial for accelerating the development time are to be provided. Such libraries may include networking, file processing and database manipulation.

Finally, and since the NLG engine is loosely-coupled with the source-to-source compiler, the language is to be internationalized by supporting more human languages such as French, German and Arabic.

ACKNOWLEDGEMENT

We thank Mrs. Henriette Skaff in the Department of Languages and Translation at AUST for her help in editing this article.

REFERENCES

[1] Brad A. Myers, *Towards More Natural Functional Programming Languages*, ICFP international conference on functional programming, 2002.
[2] Brad A. Myers, John F. Pane and Andy Ko, *Programming languages and environments*, Communications of the ACM September, 2004.
[3] Michael L. Scott, *Programming Language Pragmatics*, Second Edition, Morgan Kaufmann, 2005.
[4] Kenneth C. Louden, *Compiler Construction*, Principles and Practice, PWS, 1997.
[5] Daniel Jurafsky and James Martin, *Speech and Language Processing,* Prentice Hall, 2000.
[6] Ehud Reiter and Robert Dale, *Building Natural Language Generation Systems*, Cambridge University Press, 1999.
[7] Jeffrey Hoffer, Mary Prescott and Fred McFadden, *Modern Database Management*, 7$^{th}$ edition, Prentice Hall, 2005.
[8] J.F. Pane and B.A. Myers, *Usability Issues in the Design of Novice Programming Systems*, Carnegie Mellon University, Technical Report, 1996.
[9] Eisenberg, Resnick and Turbak, *Understanding Procedures as Objects*, Empirical Studies of Programmers: Second Workshop, Norwood, NJ, 1987.
[10] Merrill, D.C. and B.J. Reiser, *Scaffolding the Acquisition of Complex Skills with Reasoning-Congruent Learning Environments*, Proceedings of the Workshop in Graphical Representations, Reasoning, and Communication from the World Conference on Artificial Intelligence in Education. Edinburgh, Scotland, 1993.
[11] H. M. Deitel and P. J. Deitel, *C# How to Program*, 2nd edition, Prentice Hall, 1999.
[12] Anders Hejlsberg and Scott Wiltamuth, *C# Language Reference*, 2000.